\documentclass[epj]{svjour}
\usepackage{psfrag}
\usepackage{epsfig}
\usepackage{amsmath}
\usepackage{multirow}
\usepackage{rotating}
\usepackage{amssymb}
\usepackage{color}
\usepackage{axodraw}
\usepackage{bbm}
\usepackage{cite}
\newcommand{\be}{\begin{equation}}
\newcommand{\ee}{\end{equation}}
\newcommand{\ba}{\begin{array}}
\newcommand{\ea}{\end{array}}
\newcommand{\bi}{\begin{itemize}}
\newcommand{\ei}{\end{itemize}}

\newcommand{\fpzero}{{f_+^{K\pi}(0)}}
\newcommand{\fzerozero}{{f_0^{K\pi}(0)}}
\newcommand{\fpq}{{f_+^{K\pi}(q^2)}}
\newcommand{\fmq}{{f_-^{K\pi}(q^2)}}
\newcommand{\fzeroq}{{f_0^{K\pi}(q^2)}}

\begin{document}
\title{$K \to\pi$ form factors with reduced model dependence}
\subtitle{%
EDINBURGH-2010-04\\
CERN-TH/2010-024\\
SHEP-1003\\
\today
}

%Peter
\authorrunning{RBC and UKQCD Collaborations}
\titlerunning{$K\to\pi$ form factors with reduced model dependence}
\PACS{{11.15.Ha} {Lattice gauge theory} \and
      {11.30.Rd} {Chiral symmetries} \and
      {12.15.Ff} {Quark and lepton masses and mixing} \and
      {12.38.Gc} {Lattice QCD calculations} \and
      {12.39.Fe} {Chiral Lagrangians}
}
\author{P.A. Boyle$^1$, J.M. Flynn$^2$, A. J\"uttner$^3$, C. Kelly$^1$, C. Maynard$^1$, H. Pedroso de Lima$^2$, C.T. Sachrajda$^2$, J.M. Zanotti$^1$\\[1mm]
{\bf RBC-UKQCD Collaboration}\\
}%
\institute{
School of Physics and Astronomy, University of
Edinburgh,\\ Edinburgh, EH9 3JZ, UK.
\and
School of Physics and Astronomy, University of Southampton,\\
Southampton, SO17 1BJ, UK.
\and
Theory Group, Physics Department, CERN, \\CH-1211 Geneva 23,
Switzerland,
\email{juettner@mail.cern.ch}
\\}

\abstract{%
Using partially twisted boundary conditions we compute the $K\to\pi$
semi-leptonic form factors in the range of momentum transfers
$0\lesssim q^2\leq q^2_{\rm max}=(m_K-m_\pi)^2$ in lattice QCD with $N_f=2+1$ dynamical flavours.
In this way we are able to determine $\fpzero$ without any interpolation in the momentum
transfer, thus eliminating one source of systematic error. This study
confirms our earlier phenomenological ansatz for the strange quark mass dependence
of the scalar form factor. We identify and estimate potentially significant
NNLO effects in the chiral expansion that guides the extrapolation of the
data to the physical point. Our main result is
$\fpzero = 0.9599(34)(^{+31}_{-43})(14)$, where the 
first error is statistical, the second error is
due to the uncertainties in the chiral extrapolation
of the lattice data and the last error is an estimate of potential
discretisation effects.
}

\maketitle
%
%
%%%%%%%%%%%%%%%%%%%%%%%%%%%%%%%%%%%%%%%%%%%%%%%%%%%%%%%%%%%%%%%%%%%%%%%%%%%%%%
\section{Introduction}
%%%%%%%%%%%%%%%%%%%%%%%%%%%%%%%%%%%%%%%%%%%%%%%%%%%%%%%%%%%%%%%%%%%%%%%%%%%%%%
The last five years have seen tremendous progress in high precision
calculations from first principles
of the $K\to\pi$ semi-leptonic vector form factor at vanishing
momentum transfer, $\fpzero$.
The results are  combined with precise experimental measurements of $|\fpzero V_{us}$ to obtain the CKM matrix element $V_{us}$.
It has been shown in many numerical studies
 \cite{Becirevic:2004ya,Tsutsui:2005cj,
Dawson:2006qc, Brommel:2007wn,Okamoto:2004df,Boyle:2007qe,Lubicz:2009ht} that the observable $\fpzero$
can be computed with sub-percent level of precision on the lattice. 
All common efforts take advantage of the fact
that the lattice prediction is necessarily exactly unity in the $SU(3)$ limit
in which $m_\pi^2 = m_K^2$. Thus
leading lattice errors on the form factor are proportional to the mass difference and apply to the
difference of the form factor from unity
$  \fpzero - 1=\Delta f +f_2 $\,,
where $f_2$ is the analytically known $O(m_\pi^2, m_K^2, m_\eta^2)$
contribution to the form factor in $SU(3)$
chiral perturbation
theory \cite{Gasser:1984ux,Becirevic:2005py}.
The precision of current lattice computations is well represented
by the result of the RBC-UKQCD collaboration~\cite{Boyle:2007qe},
\begin{equation}\label{eqn:convres}
\Delta f = -0.0129(33)_{\rm stat.}(34)_{\rm q^2,\chi}(14)_{ a}\,.
\end{equation}
The first error is statistical, the second is due in approximately equal parts
to the chiral extrapolation and to
an interpolation in the momentum transfer
$q^2$ and the final error is due to cut-off effects.

The primary purpose of this work is to remove completely the systematic error in $\fpzero$ due
to the $q^2$-inter\-polation at the lightest quark masses. In addition we study in more detail the ansatz for the
chiral extrapolation which was used to obtain the above result (\ref{eqn:convres}).
To this end we build on our previous
work on the $q^2$-dependence of meson form factors and stochastic
quark propagators \cite{Boyle:2007wg,Boyle:2008rh,Boyle:2008yd}.
The chiral extrapolation error above includes a contribution from
using a simulated strange quark which is a little heavier than the
physical one. Here we confirm this estimate with new simulation
results which allow us to interpolate directly in the valence strange quark
mass.
We also discuss phenomenological fits based on NLO $SU(2)$ \cite{Flynn:2008tg}
and on $SU(3)$ \cite{Gasser:1984ux,Becirevic:2005py}
chiral perturbation theory
for the vector form factor  and we discuss potentially
significant NNLO effects.

Our results 
reinforce 
the RBC-UKQCD collaboration's computation
of $\fpzero$ 
which constitutes a crucial input to precision tests of the Standard Model
and for constraining its possible extensions \cite{Antonelli:2008jg}.
The results presented should be interpreted as an intermediate step towards
the extension of the RBC-UKQCD collaboration's effort towards simulations at
smaller quark masses and at a smaller lattice spacing which are under way.

After a brief discussion of the computational setup and a comparison
of our current strategy with the conventional one, we summarise and discuss
our results.
%%%%%%%%%%%%%%%%%%%%%%%%%%%%%%%%%%%%%%%%%%%%%%%%%%%%%%%%%%%%%%%%%%%%%%%%%%%%%%
\section{Strategy}
The matrix element of the vector current between initial and final
states consisting of pseudo-scalar mesons $P_i$ and $P_f$
respectively, is in general decomposed into two invariant form
factors which parameterise non-perturbative effects,
\begin{equation}\label{eqn:gen_ff}
\begin{array}{l}
\langle P_f(p_f)|V_\mu| P_i(p_i)\rangle =\\[2mm]
    \qquad\, f^{P_iP_f}_+(q^2)(p_i+p_f)_\mu +
    f^{P_iP_f}_-(q^2)(p_i-p_f)_\mu\,,
\end{array}
\end{equation}
where $q=p_f-p_i$ is the momentum transfer. For $K\to\pi$
semi-leptonic decays $V_\mu$ is the weak current $\bar{s}\gamma_\mu
u$, $P_i=K$ and $P_f=\pi$.
We also introduce the scalar form factor
\begin{equation}
 f_0^{K\pi}(q^2)=f_+^{K\pi}(q^2)+\frac{q^2}{m_K^2-m_{\pi}^2}f_-^{K\pi}(q^2)\,,
\end{equation}
 which satisfies $f_0^{K\pi}(0)=\fpzero$.
In this paper
we are primarily interested in computing $\fpq$ for $q^2=0$ but we also
present new results for other values of $q^2$. In contrast
to previous calculations of the form factor in lattice QCD
\cite{Becirevic:2004ya,Tsutsui:2005cj,
Dawson:2006qc, Brommel:2007wn,Okamoto:2004df,Boyle:2007qe,Lubicz:2009ht} we
simulate directly at $q^2=0$
by using partially twisted boundary
conditions~\cite{Sachrajda:2004mi,Bedaque:2004ax}. One
combines gauge field configurations generated with sea quarks
obeying periodic boundary conditions with valence quarks with twisted
boundary conditions \cite{Bedaque:2004kc,deDivitiis:2004kq,Sachrajda:2004mi,
Bedaque:2004ax,Tiburzi:2005hg,Flynn:2005in,Guadagnoli:2005be,
Aarts:2006wt,Tiburzi:2006px,Bunton:2006va}, i.e. the valence quarks satisfy
\begin{equation}
\psi(x_k+L)=e^{i\theta_k}\psi(x_k),\qquad(k=1,2,3)\,,
\end{equation}
where $\psi$ is either a strange quark $s$ or
it represents one of the degenerate up or down quarks $q$.
The dispersion relation for a meson in a finite volume projected onto
Fourier momentum $\vec p_{\rm FT}$ then takes the form
\cite{deDivitiis:2004kq,Flynn:2005in},
\begin{equation}\label{eqn:pi_disprel}
 E = \sqrt{m^2 + \left(\vec p_{\rm FT}+\Delta \vec \theta\right)^2},
\end{equation}
where $m$ is the meson mass and where $\Delta \vec \theta$ is the
difference between the twist angles of the two valence quarks.

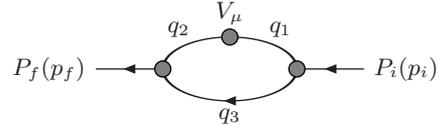
\begin{figure}
\begin{center}\begin{picture}(120,60)(-60,-30) \ArrowLine(-25,0)(-50,0)
\ArrowLine(50,0)(25,0)\Oval(0,0)(12,25)(0)
\GCirc(-25,0){3}{0.5}\GCirc(25,0){3}{0.5} \GCirc(0,12){3}{0.5}
\Text(-19,12)[b]{$q_2$}\Text(19,12)[b]{$q_1$}
\Text(0,-15)[t]{$q_3$}\Text(0,17)[b]{$V_\mu$}\Text(-54,0)[r]{$P_f(p_f)$}
\Text(54,0)[l]{$P_i(p_i)$}\ArrowLine(0.5,-12)(-0.5,-12)
\end{picture}\end{center}
\caption{Quark flow diagram for a 3pt function with initial and final states
$P_i$ and $P_f$, respectively.}
\label{fig:quarkflow}
\end{figure}

In \cite{Boyle:2007wg} we showed how this technique can be used to
extract meson transition matrix elements at arbitrary values of the momentum
transfer.
In particular, for the matrix element with the valence quark flow diagram
as in figure \ref{fig:quarkflow} with the initial and
the final meson carrying 3-momenta
$\vec p_i=\vec p_{{\rm FT},i}+\vec \theta_i/L $ and
$\vec p_f=\vec p_{{\rm FT},f}+\vec \theta_f/L $,
respectively, the momentum transfer between the initial and the
final state meson is
\begin{equation}\label{eq:mom_transfer}
q^2=(p_i-p_f)^2=\big(E_i(\vec p_i)-E_f(\vec p_f)\big)^2
        -\big(\vec{p}_i
          -\vec{p}_f\big)^2\,.
\end{equation}
In our study it will be sufficient to set the twist of the spectator
(anti-)quark $q_3$ to zero so that
it satisfies periodic boundary conditions.

As will become clear below, the matrix element in (\ref{eqn:gen_ff})
can be extracted from the time-dependence of combinations of Euclidean two- and
three-point correlation functions which can be computed straightforwardly
in lattice QCD. The two-point function is defined by
\begin{equation}
\begin{array}{rcl}
C_i(t,\vec p_i)&=&\sum_{\vec{x}}e^{i\vec{p}_i\cdot\vec{x}} \langle
\,O_i(t,\vec{x})\, O_i^\dagger(0,\vec{0})\,\rangle\\[2mm]
    &=&\frac{
            |Z_i|^2}{2E_i}
            \left(e^{-E_it}+ e^{-E_i(T-t)}\right)\, ,
\label{eq:twopt}
\end{array}
\end{equation} where $i=\pi$ or $K$, and where $O_i$ are
pseudo-scalar interpolating operators for the corresponding
mesons $O_\pi= \bar q \gamma_5 q$ and $O_K=\bar s\gamma_5 q$ and
we assume that $t$ and $T-t$ (where $T$ is the temporal extent of
the lattice) are sufficiently large that the correlation function
is dominated by the lightest state (i.e. the pion or kaon). The
constants $Z_i$ are given by
$Z_i=\langle\,P_i\,|\,O_i^\dagger(0,\vec{0})\,|\,0\,\rangle$\,. The
three-point functions are defined by
\begin{eqnarray}\label{eq:3pt}
C^{(\mu)}_{P_iP_f}(t_{i},t,t_{f},\vec p_i,\vec p_f) &=&\nonumber\\
&&\hspace{-3.7cm}=
Z_V\sum_{\vec{x}_f,\vec{x}} e^{i\vec{p}_f\cdot(\vec{x}_f-\vec{x})}
e^{i\vec{p}_i\cdot\vec{x}} \langle\, O_f(t_{f},\vec x_f)\,
V_\mu(t,\vec{x})\,O_i^\dagger(t_{i},\vec 0)\,\rangle\nonumber\\%[2mm]
&&\hspace{-3.7cm}=
Z_V
        \frac{Z_i\,Z_f}{4E_iE_f}\, \langle\,P_f(\vec{p}_f)\,|\,V_\mu(0)\,|\,
        P_i(\vec{p}_i)\,\rangle\,\nonumber\\
&&\hspace{-2.5cm}
     \times\left\{\theta(t_f-t)\,e^{-E_i(t-t_i)-E_f(t_f-t)}\ \right.\nonumber\\
&&\hspace{-2.5cm}+\,c_\mu
\left.\theta(t-t_f)\,e^{-E_i(T+t_i-t)-E_f(t-t_f)}\right\}\,,\label{eq:threept}
\end{eqnarray}
where $P_{i,f}$ is a pion or a kaon, $V_\mu$ is the
vector current with flavour quantum numbers to
allow the $P_i\to P_f$ transition and we have defined
$Z_f=\langle\, 0\,|\,O_f(0,\vec 0)|\,P_f\,\rangle$. We have introduced
the constant $c_\mu$ which is $c_0=-1$ (time-direction) and
$c_i=+1$ for $i=1,2,3$. Again we
assume that all the time intervals are sufficiently large for the
lightest hadrons to give the dominant contribution.

The vector current renormalisation factor $Z_V$ can be obtained as follows.
For illustration we take $0< t< t_f <T/2$, in which case $Z_V$ is defined by
\begin{equation}\label{eq:zv}
Z_V = \frac{\tilde C_\pi(t_f,\vec 0)}
    {C^{(B,\mu)}_{\pi\pi}(t_i,t,t_f,\vec{0},\vec{0}\,) }\,.
\end{equation}
In the numerator we use the function 
$\tilde C_\pi(t,\vec p)=
	C_\pi(t,\vec p)-
	\frac{|Z_\pi|\,^2}{2E_\pi}\,e^{-E_\pi (T-t)}$
where $Z_\pi$ and $E_{\pi}$
are determined from fits to
$C_\pi(t,\vec 0)$ and using eqn.~(\ref{eqn:pi_disprel})
(for $t_f=T/2$ it is natural instead to use
$\tilde C_\pi(t,\vec p)=\frac 12 C_\pi(t,\vec p)$ in (\ref{eq:zv})). The
superscript $B$ in the denominator indicates that we take the bare
(unrenormalised) current in the three-point function.

In the following we drop the labels $t_i$ and $t_f$ (since they are fixed)
and we combine the two- and three- point functions into the ratios
\begin{equation}\label{eq:ratios}
\begin{array}{rcl}
R^{(\mu)}_{1,\,P_iP_f}(\vec{p}_i,\vec{p}_f)&{=}&\mathcal{N}\, \sqrt{\frac
 {C^{(\mu)}_{P_iP_f}(t,\vec p_i,\vec p_f)\,
	C^{(\mu)}_{P_fP_i}(t,\vec p_f,\vec p_i)}
 {\tilde C_{P_i}(t_f,\vec p_i)\,\tilde C_{P_f}(t_f,\vec p_f)}},
 \\[4mm]
R^{(\mu)}_{3,\,P_iP_f}(\vec{p}_i,\vec{p}_f)&=&\\[2mm]
&&\hspace{-2.2cm}
\mathcal{N}\frac{C^{(\mu)}_{P_iP_f}(t,\vec p_i,\vec p_{f})}{\tilde C_{P_f}(t_f,\vec p_f)}\,
    \sqrt{
    \frac{C_{P_i}(t_f-t,\vec p_i)\,C_{P_f}(t,\vec p_f)\,\tilde C_{P_f}(t_f,\vec p_f)}
    { C_{P_f}(t_f-t,\vec p_f)\,C_{P_i}(t,\vec p_i)\,\tilde C_{P_i}(t_f,\vec
    p_i)}}\,,
\end{array}
\end{equation}
where $\mathcal{N}=4 Z_V \sqrt{E_i E_f}$ and the ratios are constructed such
that
\begin{equation}
 \begin{array}{rcl}
R^{(\mu)}_{\alpha,\,P_iP_f}(\vec{p}_i,\vec{p}_f)&=&\\[2mm]
	&&\hspace{-2.2cm}
	f^{P_iP_f}_+(q^2)({p}_i+{p}_f)_\mu+f^{P_iP_f}_-(q^2)({p}_i-{p}_f)_\mu\,,
 \end{array}
\end{equation}
for $\alpha=1,3$.
For the ratios we use the naming convention of \cite{Boyle:2007wg} but
we haven't made use of ratio $R_2$\footnote{We did not generate data
for $C_{PP}^{(\mu)}(t,\vec p,\vec p)|_{q^2=0}$ for $P=\pi,K$ from which the
 forward matrix elements $\langle P|V_\mu|P\rangle$
relevant for the construction of $R_2$ can be extracted.
}.

Once these ratios have been computed for some choices
for  $\vec p_i$ and $\vec p_f$ while keeping $q^2$ constant (of course
we are particularly interested in $q^2=0$)
the form factors $\fpq$ and $\fmq$ can be obtained as the solutions of
the corresponding system of linear equations.

%%%%%%%%%%%%%%%%%%%%%%%%%%%%%%%%%%%%%%%%%%%%%%%%%%%%%%%%%%%%%%%%%%%%%%%%%%%%%%
\section{Simulation parameters}
We simulate with $N_f=2+1$ dynamical flavours generated with the Iwasaki gauge
action \cite{Iwasaki:1985we,Iwasaki:1984cj}
at $\beta=2.13$, which corresponds to an inverse lattice spacing
$a^{-1}=1.73$
GeV ($a= 0.114(2)$fm) \cite{Allton:2007hx,Allton:2008pn},
and the domain wall fermion action \cite{Kaplan:1992bt,Shamir:1993zy} with a
residual mass of $am_{\rm res}=0.00315(2)$ \cite{Allton:2007hx}.
The simulated strange quark mass,
$am_s= 0.04$, is close to its physical value \cite{Allton:2008pn},
 and here we choose the RBC-UKQCD configurations with the bare light
sea quark mass $am_q=0.005$, which corresponds
to a pion mass of
$330$MeV \cite{Allton:2007hx,Allton:2008pn}. The calculations are
performed on a lattice volume of $24^3\times 64$ sites with the fifth
dimension having an extension of 16 lattice points.
More details can be found in \cite{Allton:2008pn}.

We distinguish three sets of correlation functions as summarised in  table
\ref{tab:simpar0}.
\begin{table}
 \centering
 \begin{tabular}{lccc}
 \hline\hline\\[-3mm]
 			& P4	&Z4PSs4&Z4PSs3\\[-0mm]
 \hline\\[-3mm]
 $am_s$			& 0.04	&0.04 &0.03\\
 \# meas			&700	&1180 &1180\\
 \# sources		&4	&8 &8\\
 \hline\hline
 \end{tabular}
 \caption{Some basic simulation parameters.}
 \label{tab:simpar0}
\end{table}
The correlation functions in set  P4 (point source, 4 positions of the source)
are identical to those
on which the RBC-UKQCD
prediction for $\fpzero$ at $am_q=0.005$ in \cite{Boyle:2007qe} is
based.
The naming convention for the remaining two data sets, $Z4PSs4$ and
$Z4PSs3$, is motivated as follows:
spin-diluted $\mathbb{Z}(2)\times \mathbb{Z}(2)$
\cite{Boyle:2008rh,Boyle:2008yd} noise source and
point sink with strange quark mass $am_s=0.04$ and $am_s=0.03$, respectively.
The data set $Z4PSs4$ with light quark mass $am_q=0.005$ corresponds to
a unitary simulation point, i.e. where the sea and valence quark masses
are the same, while set $Z4PSs3$ corresponds to a partially
quenched parameter choice.
For $Z4PSs3$ and $Z4PSs4$ we started the measurement chains for
 eight different source time-slices.
 In each case we measured on
every 40th trajectory and averaged the correlation functions over the
chains into bins of eight. The
correlation functions obtained in this way were computed with
zero Fourier momentum and the
momenta of the initial and/or final kaon/pion were induced by twisting one
of the kaon's and/or pion's valence quarks.
For each measurement we applied the full
twist along one of the spatial directions. We changed this direction
frequently as the measurements proceeded in order to reduce the
correlations. 
Our choices for the twisting angles for 
$Z4PSs4$ and $Z4PSs3$ and the corresponding values of $q^2$
are summarised in table \ref{tab:Kpi_kinematics}.
In order to obtain $q^2$=0
we make the following two choices of the twisting angles~\cite{Boyle:2007wg}:
\begin{equation}\label{eq:twists}
 \begin{array}{llcccc}
&
  |\vec{\theta}_K| =
           L\sqrt{({m_K^2+m_\pi^2 \over 2m_\pi})^2 - m_K^2}
      &\textrm{and}&\vec{\theta}_\pi=\vec{0}\,,\\[2mm]
{\rm and}&
          |\vec{\theta}_\pi| =L
          \sqrt{({m_K^2+m_\pi^2 \over 2m_K})^2 -
          m_\pi^2}&\textrm{and}
      &\vec{\theta}_K =\vec{0}\,.\\[2mm]
 \end{array}
\end{equation}
As input to these formulae we have used the estimates for the central values of
the kaon masses
$am_K=0.2990$ ($Z4PS3$) and $am_K=0.3328$ ($Z4PS4$) 
and for the pion mass $am_\pi=0.1907$ (for both datasets). 
These values had been
determined from a previous study of the gauge field ensemble considered here.
The momenta of the mesons are given by
$\vec{p}_K=\vec{\theta}_K/L$ and $\vec{p}_\pi=\vec{\theta}_\pi/L$.
In addition to the values of $\theta_\pi$ and $\theta_K$ in 
eqn.~(\ref{eq:twists}),
propagators
were generated for other values of the twisting angle. In particular, for the
kinematical situation where the kaon is 
at rest and the pion is moving  due to the additional \emph{ad hoc} twist angle
$\theta_\pi=1.600$ we determined the corresponding values for $\theta_K$ which
yield the same $q^2$ also when the pion is at rest.
The
contractions of these propagators into two- and three-point functions
allow a computation of $f_+^{K\pi}(q^2)$ for additional values
of the momentum transfer in the range from about $q^2=0$ to
$q^2_{\rm max}$ (cf. table \ref{tab:Kpi_kinematics}).

%%%%%%%%%%%%%%%%%%%%%%%%%%%%%%%%%%%%%%%%%%%%%%%%%%%%%%%%%%%%%%%%%%%%%%%%%%%%%%%%
\begin{table}
\hspace{0mm}
  \begin{tabular}{@{\hspace{1mm}}l@{\hspace{2mm}}l@{\hspace{2mm}}l@{\hspace{2mm}}l@{\hspace{2mm}}l@{\hspace{2mm}}l@{\hspace{-.5mm}}l}
  \hline\hline\\[-3mm]
    $\theta_{\pi}$&$\theta_{K}$&
        $q^2/{\rm GeV}^2$&$\fzeroq$&$\fpq$&$\fmq$\\[0mm]
   \hline\hline\\[-3mm]
   \multicolumn{6}{c}{$am_s=0.04$}\\
   \hline\hline\\[-3mm]
   0&4.68     &0.0002(2)&\multirow{2}{*}{0.9758(44)}&\multirow{2}{*}{0.9758(44)}&\multirow{2}{*}{-0.0997(93)}\\
   2.682&0    &0.0004(3)&\\
   \hline\\[-3mm]
   2.129&0    &0.0216(2)&\multirow{1}{*}{0.9898(34)}&\multirow{1}{*}{0.9975(42)}&\multirow{1}{*}{-0.081(17)}\\
   \hline\\[-3mm]
   1.600&0       &0.0381(2)&\multirow{2}{*}{1.0030(20)}&\multirow{2}{*}{1.0213(32)}&\multirow{2}{*}{-0.108(11)}\\
   0&2.792    &0.0382(2)&\\
   \hline\\[-3mm]
   0&0         &0.0607(2)
		&\multirow{1}{*}{1.0185(15)}&\multirow{1}{*}{}\\
   \hline\hline\\[-3mm]
   \multicolumn{6}{c}{$am_s=0.03$}\\
   \hline\hline\\[-3mm]
  2.682&0	&-0.0192(3)&0.9677(49) &0.9613(41)&{-0.054(14)}\\
   \hline\\[-3mm]
   0&3.337	&0.0001(5)&\multirow{2}{*}{0.9867(30)}&\multirow{2}{*}{0.9867(30)}&\multirow{2}{*}{-0.0771(77)}\\
   2.129&0	&0.0001(3)\\
   \hline\\[-3mm]
   1.600&0		&0.0149(3)&\multirow{2}{*}{0.9986(21)}&\multirow{2}{*}{1.0066(27)}&\multirow{2}{*}{-0.0852(96)}\\
   0&2.509	&0.0150(4)\\
   \hline\\[-3mm]
   0&0&0.0352(4)&\multirow{1}{*}{1.0124(10)}&\\
 \hline\hline\\
  \end{tabular}
 \caption{Table of twisting angles used in this study, together with the corresponding values of $q^2$ and
the results for the form factors.}
        \label{tab:Kpi_kinematics}
\end{table}

%%%%%%%%%%%%%%%%%%%%%%%%%%%%%%%%%%%%%%%%%%%%%%%%%%%%%%%%%%%%%%%%%%%%%%%%%%%%%%%%%
\section{Data analysis}
%%%%%%%%%%%%%%%%%%%%%%%%%%%%%%%%%%%%%%%%%%%%%%%%%%%%%%%%%%%%%%%%%%%%%%%%%%%%%%%%
The results for the two- and three-point correlation functions for
$P4$, $Z4PSs3$ and $Z4PSs4$ were analysed using the jack-knife as well as
the boot-strap procedure.
In all cases covariance matrices for the correlation functions
and ratios $R_1$ and $R_3$ were generated in order to be used for the
fits (\emph{frozen} covariance matrix). \emph{Unfreezing} the covariance matrix,
i.e. using the covariance matrix computed individually for each jack-knife
or boot-strap sample destabilised the fits. We interpret this as a
reflection of the fact that
we have an insufficient set of measurements and that
the fluctuations of the covariance matrix are therefore  large. We found
that the results we get with the frozen covariance matrix
agree within (similar) errors with the results
we would get when neglecting any correlations.

We found that for the spatial component of the vector current in the
three-point function, i.e. for $R_1^{(i)}$ and $R_3^{(i)}$, the plateaus
are of comparable quality - in the analysis we opted to use
$R_3^{(i)}$. For the time-component however
in the cases where only one of the initial
and final states carries a twist the quality of the ratios varies strongly.
Here we decided to use $R_1^{(0)}$ for the case where only the pion
carries the twist and $R_3^{(0)}$ in all other cases.

Table \ref{tab:Kpi_kinematics} summarises the kinematical points which we
analysed. The kaon masses for the full statistics of
$Z4PSs3$ and $Z4PS4$ turn out to be
$am_K=0.2987(4)$ and $am_K=0.3327(4)$,
respectively and the pion mass in both cases is
$am_\pi=0.1903(4)$\footnote{While agreeing within errors, the central values
differ slightly from those quoted in \cite{Allton:2008pn} because
 the number of measurements and the measurement techniques differ.}.
From the table we see that there are degeneracies in $q^2$, i.e.
we have data for the same $q^2$ but from three-point functions with
different kinematical parameters for the kaon and pion.
In the analysis we first compute the corresponding ratios $R_1$ and $R_3$ and
then average them as indicated by the degeneracy.
%%%%%%%%%%%%%%%%%%%%%%%%%%%%%%%%%%%%%%%%%%%%%%%%%%%%%%%%%%%%%%%%%%%%%%%%%%%%%%%%
\section{Extrapolation models}\label{sctn:exmodels}
%%%%%%%%%%%%%%%%%%%%%%%%%%%%%%%%%%%%%%%%%%%%%%%%%%%%%%%%%%%%%%%%%%%%%%%%%%%%%%%%
In a previous analysis \cite{Boyle:2007qe} we used results for the form factors
at the lattice Fourier modes
for four pion masses and one value of the (unitary)
strange quark mass which post facto turned out to be slightly heavier
than the physical one.
At each simulated light quark mass we first determined two estimates for
$\fpzero=\fzerozero$, one from an interpolation in $q^2$ with
a pole-ansatz,
\begin{equation}\label{eqn:PD}
 f_0^{K\pi}(q^2)|_{\rm pole}=\frac{f_+^{K\pi}(0)|_{\rm pole}}{1-q^2/M^2}\,,
\end{equation}
and one from an interpolation of $f_0(q^2)$
with a 2nd order polynomial, $\fpzero|_{\rm 2nd}$ (cf. table IV in
\cite{Boyle:2007qe}). We based the estimate of the systematic error due to the
phenomenological interpolation on the difference between the two results.

The final central value was determined from a global fit ansatz
incorporating pole dominance, the NLO expression for
$f_+^{K\pi}(0)$ in \cite{Gasser:1984ux},
\begin{equation}
f_+^{K\pi}(0)|_{\rm NLO}=1+f_2(f,m_\pi^2,m_K^2)
\end{equation}
($f$ is the pion decay constant), and modelling higher order contributions.
The ansatz is
\begin{equation}\label{eqn:global_fit}
 f_0^{K\pi}(q^2)=\frac{
	1+f_2+(m_K^2-m_\pi^2)^2
        \left(
        A_0+A_1(m_K^2+m_\pi^2)
        \right)}{
 	1-q^2/\left(M_0+M_1(m_K^2+m_\pi^2)\right)^2
	}\,,
\end{equation}
where we use the $N_f=2+1$ expression for $f_2$, partially quenched in the
strange quark,
as determined in \cite{Becirevic:2005py}.
Since the Kaon mass appears explicitly, after fitting (\ref{eqn:global_fit})
to our lattice data, any values for $m_\pi$ and $m_K$ can be inserted into eqn.~(15) to obtain a value for $f_0^{K\pi}(q^2)$. Hence, by inserting the physical values for $m_\pi$ and $m_K$, the strange quark mass is automatically corrected to its physical value.
The data was well described with
\begin{equation}\label{eqn:parms}
\begin{array}{llllll}
A_0&=&-0.34(9)  {\rm GeV}^{-4}\,,& A_1&=& 0.28(12) {\rm GeV}^{-6}\,,\\
M_0&=& 0.94(10) {\rm GeV}\,, &M_1&=& 0.54(18) {\rm GeV}^{-1}\,.
\end{array}
\end{equation}

\begin{figure}
 \centering
 \psfrag{qsq}[t][t][1][0]{$(aq)^2$}
 \psfrag{fplus}[l][l][1][0]{$\fpq$}
 \epsfig{scale=.4,file=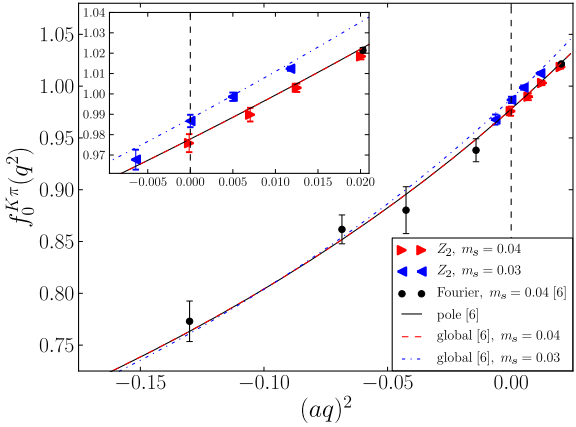}
 \caption{Summary of simulation results of $\fzerozero$. The black circles
 and the (dashed ) pole interpolation correspond to the results of
 \cite{Boyle:2007qe}
 while the results represented by the left- and right pointing arrows,
 respectively,
 correspond to the results of this work for $am_s=0.04$ and $am_s=0.03$.
 The red and blue curve represent the result from the global fit in
 \cite{Boyle:2007qe}, once for $m_K^{0.03}$ and $m_K^{0.04}$.}
 \label{fig:summary}
\end{figure}
In the meantime an expansion for $\fpzero$ in $SU(2)$ chiral perturbation
theory has been derived \cite{Flynn:2008tg}. In particular in view of the
slow convergence of $SU(3)$ chiral perturbation theory observed
for some quantities
(cf. e.g.  \cite{Allton:2008pn}) it seems useful to compare the present
extrapolation strategy
to one incorporating this new formula. Similarly to the case of $SU(3)$ chiral perturbation theory
we use the ansatz,
\begin{equation}\label{eqn:global_fitSU2}
 f_0^{K\pi}(q^2)=\frac{
	F_+(1-\frac 34 L+c_2 m_\pi^2+c_4 m_\pi^4)
        }{
 	1-q^2/\left(\tilde M_0+\tilde M_1m_\pi^2\right)^2
	}\,,
\end{equation}
where $L\equiv \frac{m_\pi^2}{16\pi^2f^2}{\rm log}
	\left(\frac{m_\pi^2}{\mu^2}\right)$ with the normalisation scale
$\mu$ and where in comparison to
the original work we have added an additional term proportional to $m_\pi^4$. Note
that the parameters in this fit ansatz depend
on the
value of the strange quark mass.\\

\noindent{\bf Reordering the chiral expansion:}
{
We do not really know how light the quarks must be for the chiral
expansion at a fixed order to represent the mass dependence of
physical quantities to a given level of precision. In principle,
(unrealistically expensive) lattice computations at very light quark
masses, perhaps even lighter than the physical ones, could answer this
question. Our present calculations however, involve masses in a regime
where NNLO terms of the $SU(3)$ expansion are non-negligible, and yet we
have insufficient data to determine these terms and all the
corresponding low-energy constants (LECs). As discussed above, we
therefore have to model the higher order contributions using an ansatz
such as that in eqn.\,(\ref{eqn:global_fit}).

We observe that the interchange symmetry $f^{K\pi}_+(q^2) = f^{\pi K}_+(q^2) $ 
is held in the $SU(3)$ expansion order by order
and also in non-per\-tur\-ba\-tive data to all orders. 
The $SU(3)$ chiral expansion in terms of the unknown,
but in principle unambiguous, LEC $f_0$ has this symmetry manifest in each 
term. However, we have the freedom to repartition terms of this expansion
between different orders: for example to use an alternative expansion
parameters $f$ differing from $f_0$ beyond leading order
$$
\fpzero = 1 + f_2(f,m_K^2,m_\pi^2) + \ldots\,.
$$
In fact, the NLO term $f_2$ is usually quoted as\linebreak
$f_2(f_\pi,m_K^2,m_\pi^2) \simeq -0.023$, with
the physical pion decay constant used in place of the unknown LEC $f_0$. 
For $SU(3)$ chiral perturbation theory 
to correspond to QCD when all
terms are summed, this must be a passive transformation on the series making
simultaneous but cancelling changes to different orders. 
There are several possible 
interpretations of the $f_\pi$ which appears in $f_2(f_\pi, m_K^2, m_\pi^2)$ 
that are consistent with QCD in the
asymptotic light mass region. 

First, we may consider $f_\pi$ entering the expression to be simply an 
estimate of the true LEC $f_0$, and a rigorous error analysis must allow for 
$f_\pi$ to vary over
a range of values.
In order that the NNLO form is unchanged, and that the 
interchange symmetry remains manifest in our ansatz, we choose this
first interpretation and vary the value of $f_0$ entering 
our expressions to estimate a systematic
error.

Second, a mass dependent expression for 
\linebreak$f_\pi(f_0,m_{ud},m_s)$ may enter $f_2$
\emph{and} the correction term 
$$ \delta_{\rm NNLO} = f_2(f_0,m_K^2,m_\pi^2)-f_2(f_\pi(f_0,m_{ud},m_s),m_K^2,m_\pi^2)\,,$$
introduced
at NNLO, passively repartitioning the series and possibly improving convergence.
 We note that this 
difference term now breaks the symmetry under $\pi\leftrightarrow K$ interchange, and
the form of NNLO terms is changed to compensate the alternate expansion parameter.

A third but less helpful interpretation would be to consider adjusting the
series to expand in terms of the physical decay constant
$f_\pi\simeq 131$~MeV. 
Passive transformation requires that 
a correction term at NNLO be introduced with the
same dependence on simulated masses
$$\delta_{\rm NNLO} = f_2(f_0,m_K^2,m_\pi^2) - f_2(f_\pi=131 {\rm MeV},m_K^2,m_\pi^2).$$ 
This simply relabels the term involving the free parameter $f_0$ as NNLO.
If $f_0$ is subsequently expanded in terms of $f_\pi$ and $m_{\pi,phys}$,
this amounts to using a phenomenological estimate for $f_0$ --- with
the associated systematic uncertainty.

Using $f_2(f=131 {\rm MeV},m_K^2,m_\pi^2)$
but \emph{failing} to adjust the forms appearing at NNLO is  
inconsistent with QCD
in the true chiral limit as it actively changes the series. 
This discussion impacts our previous analysis in which we extrapolated the defect $\Delta f$, and
did not admit variation in $f_0 \ne 131$ MeV, nor modified the form of our global fit ansatz at NNLO
to admit breaking of the mass interchange symmetry. 
While we are concerned here with the context of the $K\to\pi$
decay, the considerations are applicable to practices used in the lattice study of 
many quantities if a chiral expansion parameters differing from $f_0$ is used to improve
convergence of the truncated series.
}
\section{Results}
\noindent{\bf Cost of the simulation:}
In table \ref{tab:costs} 
we compare the cost of the simulations for the two approaches to the computation of
$\fpzero$.
\begin{table}
\begin{center}
\begin{tabular}{lc@{\hspace{0mm}}c@{\hspace{0mm}}c@{\hspace{0mm}}c@{\hspace{0mm}}c@{\hspace{0mm}}c@{\hspace{0mm}}c@{\hspace{0mm}}c@{\hspace{0mm}}c@{\hspace{0mm}}c@{\hspace{0mm}}ccccccccc}
\hline\hline\\[-4mm]
	&$am$&&$n_{\rm props}$&&$n_\theta$&&$n_{\rm src}$&&$n_{\rm config}$&&s-c&&$N_{\rm tot}$\\
\hline\\[-3mm]
$P4$	&2&$\times$&2&$\times$&1&$\times$&4&$\times$&175&$\times$&12&=&33600\\
$Z4PSs4$	&2&$\times$&2&$\times$&2&$\times$&8&$\times$&147&$\times$&4&=&37632\\
\hline\hline\\[-3mm]
\end{tabular}
\caption{Cost comparison.}
\label{tab:costs}
\end{center}
\end{table}
For $P4$, for each quark mass $am=am_q,\,am_s$  one normal and one
extended propagator (cf. definition in \cite{Boyle:2007wg}),
one twist (periodic boundary conditions), four positions of the point source on
175 configurations and 12 spin-colour inversions were necessary. Since
with point sources a Fourier transformation between the
source and the point of the current insertion can be performed almost for free,
one can directly interpolate to $q^2=0$ at no additional cost.
This is not the case when using the noise source technique as for
$Z4PSs4$ and $Z4PSs3$.
We achieved a similar precision for $f_+^{K\pi}(0)$ at approximately the same
total cost with the $Z4PS$-source type, where for each propagator of mass
$am=am_q$ or $am=am_s$
four spin-colour inversions are necessary \cite{Boyle:2008rh} for each choice of the
twist angle. We note however that
in general the quality of plateaus is significantly enhanced when using
the stochastic volume source technique.\\

\noindent{\bf ${q}^2$-dependence of the form factor:}
The data generated for this paper complements
 our previous data $P4$ by a number of new  points
for $f_0^{K\pi}(q^2)$
in the range $0\lesssim q^2\leqq q^2_{\rm max}$ for two  strange
quark masses $am_s=0.04$ (unitary) and
$am_s=0.03$ (partially quenched). The results
are illustrated in the plot in figure \ref{fig:summary} by the red/blue
right/left-pointing arrows, respectively.
The new data points for $am_s=0.04$ nicely agree with both the pole dominance
and polynomial fits
(cf. eqn.~(\ref{eqn:PD}) in \cite{Boyle:2007qe}) as can be seen in the
following comparison:
\begin{equation}
 \begin{array}{lcl}
 \multicolumn{3}{r}{\hspace{-3mm}\textrm{\bf results for $\mathbf{am_q=0.005}$, $\mathbf{am_s=0.04}$}}\\[1mm]
 f_+^{K\pi}(0)|_{\rm pole}&=&0.9774(35)\,\textrm{\cite{Boyle:2007qe}}\,,\nonumber\\
 f_+^{K\pi}(0)|_{\rm polynomial}&=&0.9749(59)\,\textrm{\cite{Boyle:2007qe}}\,,\\
 f_+^{K\pi}(0)|_{\rm this\, work}&=&0.9757(44)\,.\nonumber
 \end{array}
\end{equation}
In \cite{Boyle:2007qe} we used the spread
$f_+^{K\pi}(0)|_{\rm pole}-f_+^{K\pi}(0)|_{\rm polynomial}\approx 0.0024$
as an estimate of the
systematic due to the phenomenological $q^2$-interpolation.
As simulations move closer to the physical pion mass,
the value of $q^2_{\rm max}=(m_K-m_\pi)^2$ increases. Therefore the
interpolation to $q^2=0$, which crucially depends on the high precision which one is able to achieve
for the form factor at $q^2_{\rm max}$, will be increasingly sensitive to the ansatz one makes.
One therefore expects the systematic error due to the interpolation to increase.
We emphasise that the approach advocated here entirely removes this
uncertainty.\\

\noindent{\bf Quark mass dependence:}
Inserting the unitary and partially
quenched kaon mass which we simulated here together with
 the parameters in (\ref{eqn:parms}) into the phenomenological ansatz
(\ref{eqn:global_fit}) we can predict  the form factor that is to be expected
for $am_s=0.03$ and $am_s=0.04$ with $am_q=0.005$ as illustrated in
terms of the blue (dot-dashed) and red (dashed) 
curve in figure \ref{fig:summary}.
Both curves are nicely compatible with  the new blue and red data points, thus
confirming that the ansatz
parameterises the dependence of the form factor on a partially
quenched strange quark well.

\begin{figure}
 \centering
 \epsfig{scale=.4,file=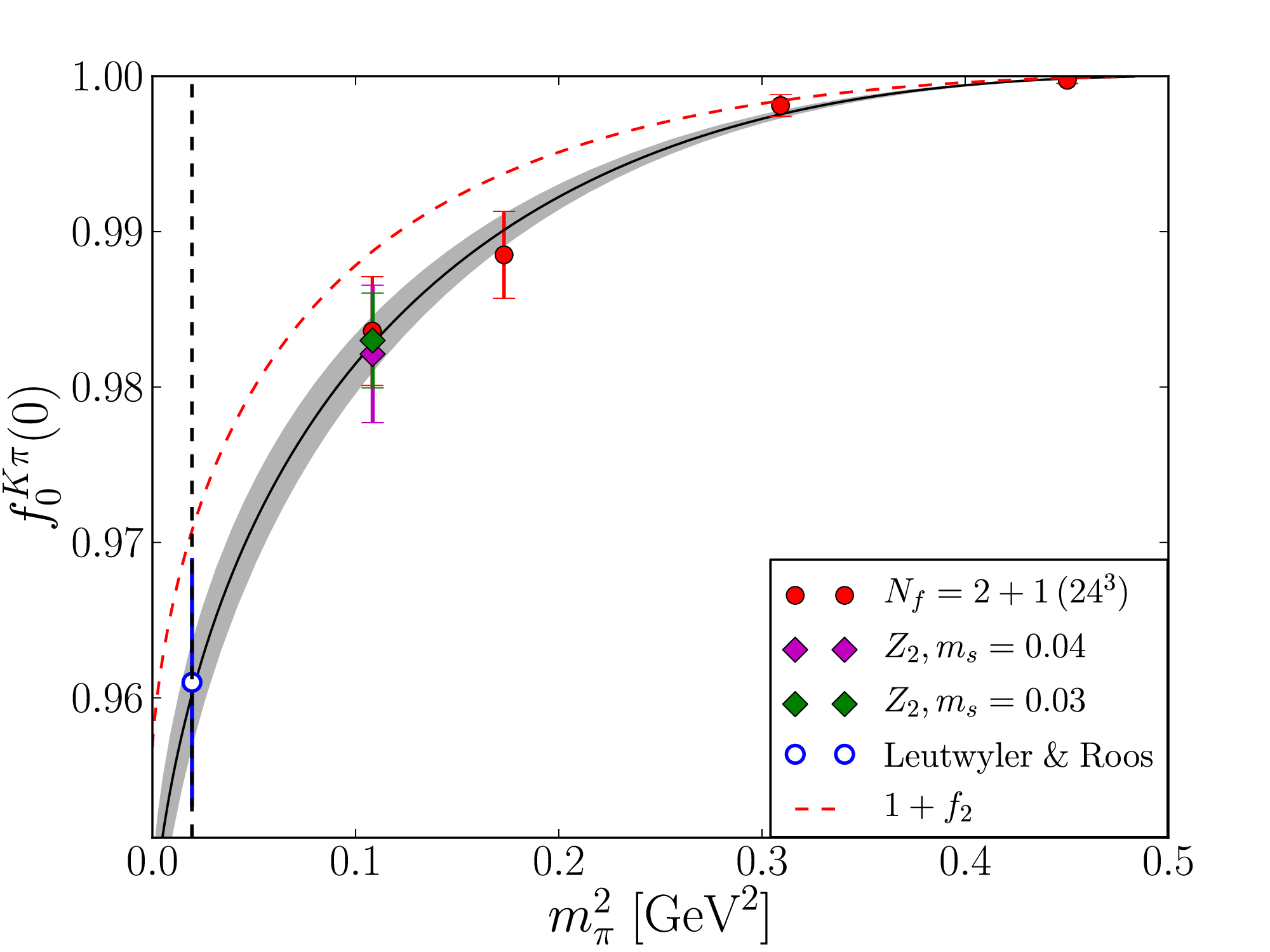}
 \caption{Result of global fit based on $SU(3)$ chiral perturbation theory
	using $f_2(115{\rm MeV},m_K,m_\pi)$. The plot also shows  a
 	comparison with  recent results
	for the form factor in chiral perturbation theory.}\label{fig:global_fit}
\end{figure}

Combining the data sets $P4$, $Z4PSs3$ and $Z4PSs4$ and carrying out the
global fit (\ref{eqn:global_fit}) we update the previous result
 $\fpzero=0.9644(33)\to\fpzero=0.9630(34)$ (statistical errors only)
at the physical point. The result of the global fit
is also illustrated in figure \ref{fig:global_fit} by the solid black line.

The chiral extrapolation of the lattice data is well constrained by the natural
\textit{hinge-point} $\fpzero|_{m_K=m_\pi}=1$. 
As can be seen in figure \ref{fig:global_fit},
our data as well as the global $SU(3)$ fit-ansatz (\ref{eqn:global_fit})
nicely approach this point for $m_\pi\to m_K$.
In contrast, in
$SU(2)$ chiral perturbation theory one expands the form factor around
vanishing pion mass at a fixed strange quark mass \cite{Flynn:2008tg}
(in fact, all strange quark mass dependence resides in the low energy
constants). The limit $\fpzero|_{m_s=m_q}=1$ is not naturally implemented in
this expansion.
Given  our experience with $SU(2)$ fits to other pion and kaon observables
in \cite{Allton:2008pn} such an expansion describes the lattice data reliably
only below
$m_\pi\approx 400$MeV. In contrast to our study in
\cite{Allton:2008pn} here we only have data for
two values of the pion mass below this cut-off and
extrapolations are therefore not well constrained. Alternatively one can
include data at heavier pion masses. However, fits of acceptable quality
can then only be obtained after adding an extra term ($\propto c_4$) to the
expression in \cite{Flynn:2008tg}. Given these considerations,
at this stage we refrain
from presenting fit results based on $SU(2)$ chiral perturbation theory.\\

\noindent{\bf Estimates of systematic errors:}
The new data presented here
confirms the ansatz for the $q^2$-interpolation for the smallest mass used in ref.\,\cite{Boyle:2007qe}, i.e.
$am_q=0.005$. Since $q^2_{\rm max}$ increases as $m_q$ decreases, it is at this mass that $q^2_{\rm max}$ is the largest (and therefore furthest away from $q^2=0$) and hence the interpolation is the least constrained.
We are therefore confident that the pole ansatz previously used in fits to
our data \cite{Boyle:2007qe} describes the form factor
data well also for all the other simulation parameters where  $q^2_{\rm max}$ is
closer to the origin.
The systematic error due to the interpolation can be safely removed
from our final result.

\begin{figure}
\begin{center}
\epsfig{scale=.3,angle=-90,file=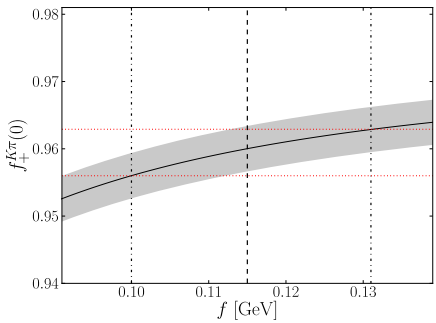}
\end{center}
\caption{
Illustration of the dependence of the fit result (with the ansatz  in eqn. 
(\ref{eqn:global_fit})) on the choice of the decay constant. 
The  horizontal red
lines indicate our estimate of the resulting systematic uncertainty.
}\label{fig:f2_f}
\end{figure}

As discussed in section \ref{sctn:exmodels}, 
a potential source of systematic error which
 to our knowledge
has not been taken into account systematically in any previous computation
of $\fpzero$ is the choice of the decay constant entering in the
$SU(3)$ NLO prediction for the form factor.
Lacking a precise value of the decay constant in the chiral limit we 
repeated the global fit for the three choices 
$f=100$, $115$, $131$\,MeV and found for the central values of the form-factor $\fpzero=0.9556,\,0.9599,\,0.9630$, respectively. In each case the fit was of very good 
quality.
This is
quite a sizeable variation in the central value which is illustrated in 
figure \ref{fig:f2_f}.
We found that the choice
of decay constant particularly changes the slope of $\fpzero$ with respect
to $m_\pi^2$ in the region of small pion masses where we do not have data.
In order to study the behaviour at NNLO more systematically, one can use the 
FORTRAN computer code
written by Bijnens\footnote{
the programs for $f_+(q^2)$ and $f_-(q^2)$ used in Ref. \cite{Bijnens:2003uy}
are available on request from Johan Bijnens} that can provide  $SU(3)$
NNLO terms in terms of numerical integration routines
(see also \cite{Bernard:2009ds} for
form factor fits based on this code). Our
experience from using the code is that for our limited set of
lattice data points there are too many free
parameters (low energy
constants from the $O(p^4)$ and the $O(p^6)$ chiral Lagrangian)
to be able to carry out reliable fits. 
Lacking a better analytical understanding of NNLO effects
we prefer as the central value the one corresponding
to $f=115$\,MeV. As an estimate for the uncertainty in the chiral extrapolation
we use the interval defined by the result for $\fpzero$ as obtained when using
$f=100$MeV and $f=131$\,MeV, respectively. Our result is therefore,
\begin{equation}
\begin{array}{rcl}
\fpzero&=&0.9599(34)(^{+31}_{-43})(14)\,,
\end{array}
\end{equation}
where the first line the first error is statistical, the second is due to the
chiral extrapolation
and the third error is an estimate of lattice cut-off effects for which
we stay with the previous estimate in \cite{Boyle:2007qe}.
We note that the quoted uncertainty due to the chiral extrapolation 
covers the central value of the result which one obtains when extrapolating 
instead with 
$f_2(f_\pi(m_{ud},m_s),m_K^2,m_\pi^2)$ in (15), i.e. with the decay constant 
as input that
corresponds to each of our simulations points (c.f. \cite{Allton:2008pn}).
Adding all errors in quadrature we obtain
\begin{equation}
\begin{array}{rcl}
\fpzero&=&0.960(^{+5}_{-6})\,.
\end{array}
\end{equation}
We believe that the systematic error due to the chiral extrapolation discussed
above is conservative but still a more rigid statement would be desirable.
To this end a better understanding of the NNLO terms in the chiral expansion and
additional simulation points at smaller pion masses are mandatory.

%%%%%%%%%%%%%%%%%%%%%%%%%%%%%%%%%%%%%%%%%%%%%%%%%%%%%%%%%%%%%%%%%%%%%%%%%%%%%%%%
\section{Summary and Outlook}
In this paper we present new data for the kaon semi-leptonic decay form factors computed in lattice QCD with $N_f=2+1$ flavours of dynamical
fermions. Using (partially) twisted boundary conditions we were able to directly simulate at the phenomenologically relevant kinematical point $q^2=0$,
at
a very small additional computational cost while at the same time making the
computation independent of any phenomenological ansatz for the interpolation
in the momentum transfer. In this way one significant systematic
uncertainty in the computation of the form factors has been successfully
removed.\\
We have reconsidered the estimates of the systematic uncertainties of our
previous calculation presented in \cite{Boyle:2007qe} and we present numerical
evidence that our ansatz for the strange quark mass dependence of the form
factor is under control by providing data at an additional partially quenched
simulation point for the strange quark mass. \\
Currently, chiral extrapolations of lattice results for
the kaon semi-leptonic form factor are based on NLO chiral perturbation theory.
We show that ambiguities in the parametrisation of the NLO expression can
lead to additional systematic effects which we include into our revised estimate
of the systematic uncertainties.
This ambiguity also applies to any other lattice computation of $\fpzero$.\\
We want to stress that the interpretation of lattice data for the
$K\to\pi$ form factors would profit from the availability of their expressions
at NNLO in chiral perturbation theory in a more transparent form.\\
The RBC-UKQCD collaboration is currently extending the set of data points
presented here by simulations at lighter  pion masses and at the same
time at a finer lattice spacing. A combined analysis of all data is the
next step in RBC-UKQCD's program of a precise computation of the
$K\to\pi$ form factors in $N_f=2+1$ flavour lattice QCD.\\

%%%%%%%%%%%%%%%%%%%%%%%%%%%%%%%%%%%%%%%%%%%%%%%%%%%%%%%%%%%%%%%%%%%%%%%%%%%%%%%%
{\bf Acknowledgements:}
We thank the members of the RBC and UKQCD Collaborations. We are grateful
to Johan Bijnens for providing us with his FORTRAN code and  
to the Engineering and Physical Sciences Research Council (EPSRC) for a 
substantial allocation of time on HECToR under the Early User initiative. 
JMF, HPdL and CTS acknowledge support from STFC Grant ST/G000557/1 and EU contract MRTN-CT-2006-
035482 (Flavianet). JZ is supported by STFC grant ST/F009658/1. PAB is supported by a RCUK fellowship. The computations reported here were carried out on the QCDOC supercomputer at the University of Edinburgh.

\bibliographystyle{JHEP}
\bibliography{kl3_systematics}
\end{document}